\documentclass[aip,pof,reprint,onecolumn,groupedaddress]{revtex4-1}
\usepackage{epsfig,graphics,amssymb,amsmath,subeqnarray,color}
\usepackage[sfdefault=cmbr]{isomath}

\usepackage{bm}
\def\boldsymbol{\bm}

\def\para{\parallel}

\def \t{\tensorsym}
\def \lb{\left}
\def \rb{\right}

\def \d{\,\text{d}}

\def \eps{\epsilon}

\def \etah{\hat{\eta}}
\def \bgamma{\boldsymbol{\gamma}}

\def \bgammad{\dot{\boldsymbol{\gamma}}}
\def \bgammadh{\hat{{\dot{\bgamma}}}}

\def \bnabla{\boldsymbol{\nabla}}

\def \bsigma{\boldsymbol{\sigma}}

\def \bsigmah{\hat{\boldsymbol{\sigma}}}

\def \btau{\boldsymbol{\tau}}

\def \btauh{\hat{\boldsymbol{\tau}}}

\def \bOmega{\mathbf{\Omega}}
\def \bOmegah{\hat{\mathbf{\Omega}}}

\def \bXi{\mathbf{\Xi}}

\def \De{\text{De}}

\def \Re{\text{Re}}

\def \bF{\mathbf{F}}

\def \bG{\mathbf{G}}

\def \bI{\mathbf{I}}

\def \bL{\mathbf{L}}

\def \bN{\mathbf{N}}

\def \bR{\mathbf{R}}

\def \bU{\mathbf{U}}

\def \bRh{\hat{\mathbf{R}}}

\def \bUh{\hat{\mathbf{U}}}

\def \bc{\mathbf{c}}

\def \bk{\mathbf{k}}

\def \bn{\mathbf{n}}

\def \br{\mathbf{r}}

\def \bu{\mathbf{u}}

\def \bx{\mathbf{x}}

\def \bzero{\mathbf{0}}

\def \buh{\hat{\bu}}

\def \fA{\mathcal{A}}
\def \fB{\mathcal{B}}

\def \fO{\mathcal{O}}

\def \tzero{\mathsf{\t 0}}

\def \tF{\mathsf{\t F}}

\def \tU{\mathsf{\t U}}

\def \tFh{\mathsf{\t{\hat{F}}}}
\def \tGh{\mathsf{\t{\hat{G}}}}

\def \tRh{\mathsf{\t{\hat{R}}}}

\def \tTh{\mathsf{\t{\hat{T}}}}
\def \tUh{\mathsf{\t{\hat{U}}}}

\begin{document}

\title{The effect of gait on swimming in viscoelastic fluids}
\author{Gwynn J. Elfring}\email{gelfring@mech.ubc.ca}
\affiliation{
Department of Mechanical Engineering, 
University of British Columbia}
\author{Gaurav Goyal}
\affiliation{
Department of Mechanical Engineering, 
University of British Columbia}
\date{\today}
\begin{abstract}
In this paper, we give formulas for the swimming of simplified two-dimensional bodies in complex fluids using the reciprocal theorem. By way of these formulas we calculate the swimming velocity due to small-amplitude deformations on the simplest of these bodies, a two-dimensional sheet, to explore general conditions on the swimming gait under which the sheet may move faster, or slower, in a viscoelastic fluid compared to a Newtonian fluid. We show that in general, for small amplitude deformations, a speed increase can only be realized by multiple deformation modes in contrast to slip flows. Additionally, we show that a change in swimming speed is directly due to a change in thrust generated by the swimmer.
\end{abstract}

\maketitle

\section{Introduction}
The locomotion of microscopic cells through viscous fluids is common in many areas of biology, from microbes searching for food \citep{fenchel02,stocker08} or causing diseases \citep{josenhans02}, to sperm cells in mammalian reproduction \citep{suarez06}. The fluid forces generated by a deforming body at low Reynolds numbers are arguably well understood for Newtonian fluids \citep{brennen77, lauga09b}, but insight is far more limited when considering bio-locomotion through fluids that display non-Newtonian rheology \citep{lauga15}.

Many biological fluids like blood, mucus, saliva and synovial fluid, demonstrate viscoelasticity and shear-thinning viscosity \citep{merrill69, hwang69, fung13}. A viscoelastic fluid retains a memory of its flow history, while a shear-thinning fluid experiences a decrease in apparent viscosity with applied strain-rate and it is in such fluid environments that many microorganisms swim; \textit{Helicobacter pyroli} in gastric mucus \citep{celli09}, and spermatozoa wading through cervical mammalian mucus \citep{fauci06} are common examples. Swimming in complex fluids can be substantially different than in Newtonian fluids, for example, locomotion in complex fluids is not constrained by the scallop theorem \citep{purcell77} meaning reciprocal swimming strokes, which produce no net motion whatsoever in Newtonian fluids, can propel a swimmer in complex fluids \citep{lauga09,keim12,qiu14}. 

Several recent articles have investigated changes in swimming kinematics due to nonlinear viscoelasticity theoretically \citep{lauga07, fu07, lauga09, fu09, balmforth10, pak12, curtis13, lauga14, yazdi14, riley14, riley15,li15, yazdi15, bohme15, decorato15}, numerically \citep{teran10, zhu12, spagnolie13, thomases14, li14, li15b} and experimentally \citep{liu11, shen11, dasgupta13, espinosa13, gagnon14, qin15}, while comprehensive reviews have summarized key findings in the theory \cite{elfring15b}, and experiments \cite{sznitman15}, of biolocomotion in complex fluids. The picture that emerges from recent studies on the effects of viscoelastic fluids, is that whether a swimmer goes faster or slower depends on the type of gait \cite{teran10,thomases14} or the amplitude of the gait \cite{liu11, spagnolie13}. Several studies have used the simplified models such as an infinite sheet \cite{riley14,elfring15b,riley15} or an infinite helix \cite{li15} undergoing general small-amplitude deformations to shed light on how a particular gait (or which) may lead to faster (or slower) swimming in the presence of a nonlinearly viscoelastic fluid. It is in this vein that we proceed here, extending the recent results of Riley and Lauga \cite{riley15} to a broader class of boundary conditions on flat body. We discuss in detail why all individual deformation modes lead to slower swimming and how one may obtain faster swimming by determining the nature of the nonlinear response of the fluid. We also show that a prescribed slip velocity can lead to an entirely different viscoelastic response compared to a prescribed deformation. To do this, we first derive integral theorems for the locomotion of two-dimensional bodies in complex fluids \cite{lauga14}, extending to non-Newtonian fluids recent results on the use of the reciprocal theorem for the swimming of two-dimensional bodies in Newtonian fluids \cite{elfring15}. Through the use of these integral theorems one can show that slower swimming is directly the result of a reduction of thrust in a viscoelastic fluid for any single (small amplitude) deformation mode. 

\section{Swimmer motion}
The motion of the surface, $S(t)$, of a shape changing swimmer may be decomposed into translation, rotation and deformation as follows
\begin{align}\label{swimmerbc}
\bu(\bx_S)=\bU+\bOmega\times\br_S+\bu^S,
\end{align}
where $\bx_S\in S(t)$. The first two terms represent rigid-body motion, $\bU$ is the instantaneous translation, $\bOmega$ is the angular velocity, while the third term $\bu^S$ is deformation due to a swimming gait \cite{felderhof12,elfring15b}. One often defines $\br_S$ from the center of mass (extensive discussion is provided elsewhere \cite{yariv06, ishimoto12, felderhof12, elfring15}) and periodic deformations are written as deviations from a (simple) reference surface $S_0$ in a body-fixed frame as
\begin{align}
\br_S = \br_0+\Delta\br(\br_0,t),
\end{align}
and then $\bu^S = \partial \Delta \br/\partial t$ (see figure \ref{swimmerbc}).

\begin{figure}[t]
\centering
\includegraphics[width=0.5\textwidth]{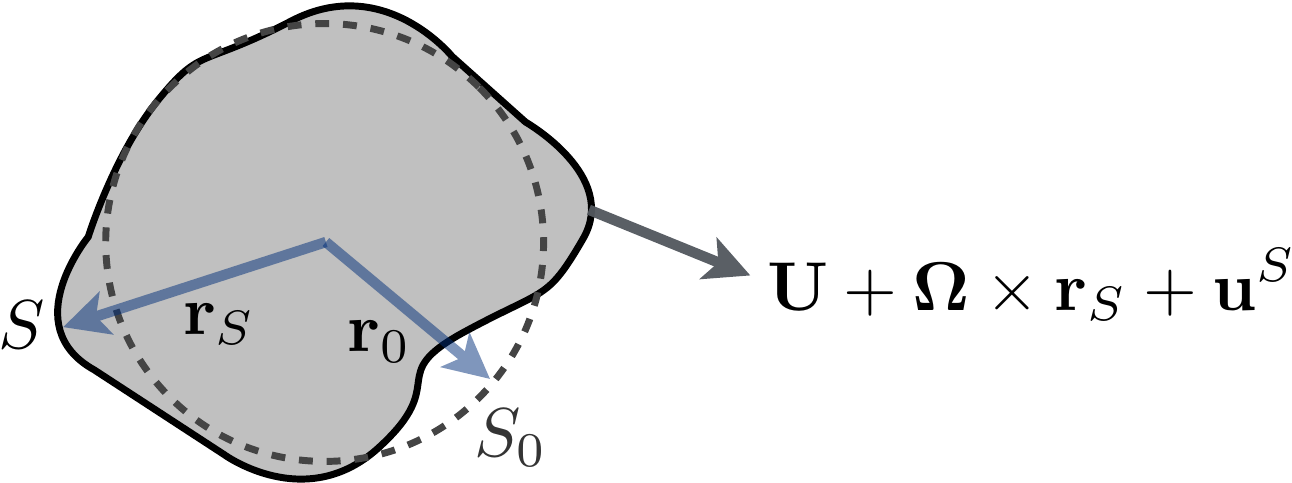}
\caption{Schematic representation of a general swimmer. A swimmer is defined as body with whose surface deforms in time thereby effecting an instantaneous rigid-body translation, $\bU$, and rotation, $\bOmega$.}
\label{swimmer}
\end{figure}

We describe here the locomotion of microorganisms small enough that the Reynolds number of the flows generated may be taken to be zero
\begin{align}
\bnabla\cdot\bsigma = -\bnabla p+\bnabla\cdot\btau = \bzero,
\end{align}
where $\bsigma=-p\bI+\btau$ is the stress tensor of a fluid with velocity, pressure and deviatoric stress fields $\bu$, $p$, $\btau$, respectively. Additionally the bodies themselves are considered instantaneously force and torque free,
\begin{align}
\bF &= \int_S \bn\cdot\bsigma \d S =\bzero,\label{force}\\
\bL &= \int_S \br_S\times(\bn\cdot\bsigma) \d S =\bzero\label{torque},
\end{align}
where the surface $S$ is a function of time and the normal to the surface $\bn$ points into the fluid. For compactness we introduce six dimensional vectors that contain both force and torque $\tF = [\bF \ \ \bL]^\top$ and translation and rotation $\tU = [\bU \ \ \bOmega]^\top$.

\section{The complex reciprocal theorem}\index{Reciprocal theorem}
Stone and Samuel showed that determining the rigid-body motion of a deforming swimmer \cite{stone96}, $\bU$ and $\bOmega$, may be greatly simplified by appealing to the Lorentz reciprocal theorem \cite{happel83}. They showed that by using the solution to the auxiliary rigid-body resistance problem one may solve for the swimming kinematics of a deforming body without resolving the flow field it generates. Lauga later extended the use of the reciprocal theorem for swimming to non-Newtonian fluids \cite{lauga09} and these ideas were then subsequently further developed \cite{lauga14,elfring15b} and integral theorems have subsequently been used in a number of recent theoretical studies of locomotion in complex fluids \cite{pak12,yazdi14,yazdi15,bohme15,datt15}. We present here integral theory for a general non-Newtonian fluid in the formalism of Elfring and Lauga \cite{elfring15b} before showing its application to simple bodies.

For a force- and torque-free swimmer of surface $S$ in a non-Newtonian fluid we denote the velocity field and its associated stress tensor by $\bu$ and $\bsigma$, while for a body of the same instantaneous shape subject to (an arbitrary) rigid-body translation and rotation $\bUh$ and $\bOmegah$ in a Newtonian fluid ($\btauh=\etah\bgammadh$) the velocity field and its associated stress tensor are denoted by $\buh$ and $\bsigmah$.  Each fluid is in mechanical equilibrium and hence the mixed products $\buh\cdot(\bnabla\cdot\bsigma)=\bu\cdot(\bnabla\cdot\bsigmah)=\bzero$, then by integrating over the volume of fluid $V(t)$ external to the surface $S(t)$ with normal $\bn$ (positive into the fluid) and invoking the divergence theorem one obtains
\begin{align}\label{11}
\int_S \bn\cdot\bsigma\cdot\buh\d S +\int_V \bsigma:\bnabla\buh\d V=\int_S \bn\cdot\bsigmah\cdot\bu\d S+\int_V \bsigmah:\bnabla\bu\d V=0.
\end{align}
The first term on the left-hand side of Eq.~\eqref{11} is zero because the swimmer is force- and torque-free, and hence the second term on the left-hand side is also zero, by construction. The first term on the right-hand side of Eq.~\eqref{11} may be expanded by using the boundary motion on $S$ in \eqref{swimmerbc} and so, assuming the fluids are incompressible, we have
\begin{align}
\tFh\cdot\tU&=-\int_S \bn\cdot\bsigmah\cdot\bu^S\d S-\etah\int_V \bgammadh:\bnabla\bu\d V,\label{int1}\\
0 &= \int_V \btau:\bnabla\buh\d V.\label{int2}
\end{align}
Here $\tFh$ represents the force and torque resulting from the rigid-body motion of $S$.

Due to the linearity of the Stokes equations we may write $\buh = \tGh\cdot\tUh$, $\bsigmah = \tTh\cdot\tUh$ while $\tFh = -\tRh\cdot\tUh$. We also assume a simplistic decomposition of the constitutive relation into linear and nonlinear components $\btau = \eta\bgammad+  \bN(\bu,\btau)$, but, as shown by Lauga\cite{lauga14}, we expect such a form to arise when solving the problem perturbatively, either expanding in a weakly non-Newtonian limit or a small deformation limit, in either case, the effects of nonlinearities are small compared to the leading order Newtonian behavior. Substituting into \eqref{int1} and utilizing \eqref{int2} while discarding the arbitrary vector $\tUh$ yields
\begin{align}\label{motiongeneral}
\tU=\tRh^{-1}\cdot\lb[\int_S \bu^S\cdot(\bn\cdot\tTh)\d S -  \frac{\etah}{\eta}\int_V \bN:\bnabla\tGh\d V\rb].
\end{align}
The first term in the brackets represents thrust generated in a Newtonian fluid (as we show below) whereas the volume integral only contributes if the fluid in the swimming problem is non-Newtonian and hence is a measure of the modification of the swimming dynamics due to the presence of non-Newtonian stresses.

One may also obtain integral statements for drag and thrust in complex fluids. In the drag problem, the body is simply undergoing rigid body motion, namely $\bu_D(\bx^S) = \bU + \bOmega\times\br_S$, whereas in the thrust problem the body is deforming but otherwise held fixed in place $\bu_T(\bx^S) = \bu^S$. Due to the linearity of the Stokes equations, these two flow fields sum to give the flow field due to swimming in a Newtonian fluid, but in a complex fluid this is in general not the case due to the nonlinearity of the constitutive relation. Following the approach above we can show that the drag on a body of shape $S$ under rigid body motion
\begin{align}\label{eq:drag}
\tF_D = -\frac{\eta}{\etah}\tRh\cdot\tU -  \int_V \bN_D:\bnabla\tGh\d V.
\end{align}
Whereas the thrust generated by a deforming body with $\bu_T(\bx^S) = \bu^S$, is
\begin{align}\label{eq:thrust}
\tF_T=\frac{\eta}{\etah}\int_S \bu^S\cdot(\bn\cdot\tTh)\d S-\int_V \bN_T:\bnabla\tGh\d V.
\end{align}
The nonlinear components, $\bN_D=\bN(\bu_D,\btau_D)$ and $\bN_T=\bN(\bu_T,\btau_T)$, depend on the respective drag and thrust fields. As we can see, taking $\tF_D+\tF_T=\tzero$ does not, in general, lead to \eqref{motiongeneral}, unlike for a Newtonian fluid, because $\bN \ne \bN_T+\bN_D$.

We note that the formulas derived here are independent of the choice of viscosity, $\etah$, in the auxiliary Newtonian problem. In particular, there is no requirement that $\etah=\eta$ although such a choice (when sensible) will simplify the appearance of the formulas.

\subsection{Small amplitude deformations}
A deforming body will have a time dependent shape, $S(t)$, but if the amplitude of the deformation is small, then we can, through Taylor series recast the problem onto a simpler, static reference surface, $S_0$ with boundary velocity $\bu^{S_0}$ \cite{felderhof94a,felderhof94b}. The velocity field is expanded
\begin{align}\label{velonbody}
\bu(\br_S)&=\bu(\bx_0)+\Delta\br\cdot\lb.\bnabla\bu\rb|_{\bx_0}+\fO(|\Delta\br|^2),
\end{align}
where $\Delta\br = \br_S-\br_0$ and $\bx_0\in S_0$. Rearranging we obtain the boundary condition on $S_0$,
\begin{align}\label{bcsmall}
\bu(\bx_0) = \bU +\bOmega\times\br_0+\bu^{S_0},
\end{align}
where
\begin{align}\label{speedS0}
\bu^{S_0}&=\frac{\partial\Delta\br}{\partial t}+ \bOmega\times\Delta\br-\Delta\br\cdot\lb.\bnabla\bu\rb|_{\bx_0}+\fO(|\Delta\br|^2).
\end{align}

Note that the swimming gait on $S_0$, $\bu^{S_0}$, depends on gradients of the (unknown) flow field $\bu$ and the rotation rate of the swimmer $\bOmega$. However if we take $\Delta \br = \epsilon\br_1$ where $\epsilon \ll 1$, is a dimensionless measure of gait amplitude, then expand the velocity field in a regular perturbation series, $\bu=\sum_m \epsilon^m\bu_m$, with boundary condition $\bu^{S_0}(\epsilon) = \epsilon\bu_1^{S_0}+\epsilon^2\bu_2^{S_0}+...$ where
\begin{align}
\bu_1^{S_0}&=\frac{\partial\br_1}{\partial t},\label{bcformula1}\\
\bu_2^{S_0}&=\bOmega_1\times\br_1-\br_1\cdot\lb.\bnabla\bu_1\rb|_{\bx_0},\label{bcformula2}
\end{align}
we obtain boundary conditions for the velocity field on $S_0$, which are known order-by-order. Furthermore, upon expanding all fields in $\epsilon$, we see that the leading order effect of the nonlinearity in the constitutive equation enters, at most, at quadratic order $\bN=\eps^2\bN_2[\bu_1,\btau_1]$. We proceed by applying the reciprocal theorem in \eqref{motiongeneral} directly on $S_0$, which is permissible because the stress in the fluid between $S$ and $S_0$ is divergence free\cite{elfring15b}, to obtain
\begin{align}
\tU&=\epsilon\tRh^{-1}\cdot\int_{S_0} \bu_1^{S_0}\cdot(\bn\cdot\tTh)\d S+\epsilon^2\tRh^{-1}\cdot\lb[\int_{S_0} \bu_2^{S_0}\cdot(\bn\cdot\tTh)\d S -  \frac{\etah}{\eta}\int_{V_0} \bN_2:\bnabla\tGh\d V\rb]+\fO(\epsilon^3).
\end{align}
The first term typically does not contribute to steady-state swimming as for a periodic gait we have $\overline{\bu_1^{S_0}}=\bzero$ (where here an overbar denotes a time-average over a period $2\pi/\omega$). If the boundary conditions have $\epsilon\rightarrow-\epsilon$ symmetry then we should expect $\lb<\bu_1^{S_0}\rb>=\bzero$ (where the angle brackets denote a surface average over $S_0$), which leads to zero net velocity to leading order for simple bodies as we will see. When there is no net motion of the body at leading order, $\tU_1 = \tzero$, then the boundary conditions in the swimming problem are precisely equal to a thrust problem where the body is held fixed to leading order, $\bu_1(\bx_0)=\bu_1^{S_0}$. In this case it follows that the first viscoelastic correction, is equal in the swimming and thrust problems, $\bN_2=\bN_{T,2}$, in other words the change in swimming speed is due entirely to the modification of the thrust by the complex rheology. As a result, to leading order, Newtonian drag (but with viscosity $\eta$) balances the thrust generated in a viscoelastic fluid
\begin{align}\label{Usmall}
\tU = \frac{\etah}{\eta}\tRh^{-1}\cdot\tF_T,
\end{align}
where the thrust is given by the reciprocal theorem \eqref{eq:thrust}. As we shall show below, for a generalized swimming sheet in a viscoelastic fluid, the thrust for any single temporal deformation mode is diminished thereby causing a reduction of swimming speed.

\section{Viscoelastic fluid relations}
In this work we consider a general constitutive relation for polymeric fluids given by
\begin{align}
\fA_j\btau^{(j)}=\eta_0^{(j)}\fB_j\bgammad+\bN^{(j)}(\bu,\btau^{(j)}),\label{constrelgen}
\end{align}
where the deviatoric stress tensor written as a sum of $j$ relaxation modes, $\btau = \sum_j \btau^{(j)}$, $\eta_0^{(j)}$ is the zero-shear-rate viscosity for the $j$-th mode, $\fA_j$ and $\fB_j$ are linear operators in time and $\bN^{(j)}$ is a symmetric tensor that depends nonlinearly on the velocity and stress and represents the transport and stretching of the polymeric microstructure by the flow \cite{bird87a,lauga09}.

We shall neglect here the influence of a particular initial stress state in the fluid, suitable when determining the steady swimming speed of a microorganism, in which case we may write the flow and stress fields as time periodic, expressed generally as
\begin{align}
\bu = \sum_p \bu^{(p)}e^{-ip\omega t}.
\end{align}
Upon summing over all relaxation modes $(j)$ we may write the constitutive relationship for each temporal mode $(p)$ as
\begin{align}\label{constrel}
\btau^{(p)}=\eta^*(p)\bgammad^{(p)}+\bN^{(p)},
\end{align}
where $\eta^*(p)=\sum_j\eta_0^{(j)}[B_j(p)/A_j(p)]$ and $\bN^{(p)}=\sum_j[1+A_j(p)]^{-1}\bN^{(j,p)}$ where
 $A_j(p)$ and $B_j(p)$ are the characteristic polynomials of the differential operators (i.e $e^{ip\omega t}\fA_j[e^{-ip\omega t}]$). For each Fourier mode we thus have a linear response with complex viscosity, $\eta^*(p)$, and a nonlinear term. As an example, the Oldroyd-B equation, which has a single relaxation mode, yields $\eta^*(p) = \eta_0(1-ip \omega \lambda_2)/(1-ip\omega\lambda_1)$ where $\lambda_1$ is the relaxation time and $\lambda_2$ is the retardation time.

Subsitution of Eq.~\eqref{constrel} into Eq.~\eqref{motiongeneral} and recasting onto a static shape $S_0$ we obtain a spectral decomposition of the swimming velocity in a non-Newtonian fluid
\begin{align}\label{mgrecast}
\tU^{(p)}=\tRh^{-1}\cdot\bigg[&\int_{S_0} \bu^{S_0,(p)}\cdot(\bn\cdot\tTh)\d S-\frac{\etah}{\eta^*(p)}\int_{V_0} \bN^{(p)}:\bnabla\tGh\d V\bigg].
\end{align}
We are often interested in only the zeroth mode, or mean swimming velocity, $\tU^{(0)}=\overline\tU$. In either case, we see that the non-Newtonian contribution arises directly as a result of the tensor $\bN$ and therefore, linearly viscoelastic fluids yield precisely the same swimming velocity as Newtonian fluids for a given prescribed gait.

\section{Model Swimmers}
It is typical, for analytical tractability, for $S_0$ to align with a coordinate system, in particular, spherical, cylindrical and planar swimmers. For two-dimensional swimmers the resistance problem is ill-posed, but as shown previously for Newtonian fluids \cite{elfring15}, use of the reciprocal theorem for swimming is still valid in two dimensions. Here, we provide simplified reciprocal theorem formulas for spherical, cylindrical and planar swimmers in complex fluids.

\subsection{Spherical swimmers}
If the body is a sphere of radius $a$, then the rigid body problem leads to resistance tensors $\bRh_{FU}=6\pi\etah a\bI$, $\bRh_{L\Omega}=8\pi\etah a^3\bI$, $\bRh_{F\Omega}=\bzero$ and $\bRh_{LU}=\bzero$, while traction on the surface gives $\bn\cdot\tTh = -\frac{3\etah}{2a}[\bI \ \ 2\bXi]$ where $\Xi_{ij}=\epsilon_{ijk}r_k$. With these relations, the swimming speed for a sphere
\begin{align}
\tU &= -\frac{1}{4\pi a^2}\int_{S_0}
\begin{bmatrix}
\bI\\
\frac{3}{2a^2}\bXi^\top
\end{bmatrix}\cdot\bu^{S_0}\d S-\frac{\etah}{\eta}\int_{V_0} \bN:\bnabla\tGh\cdot\tRh^{-1}\d V,
\end{align}
where for a sphere
\begin{align}
\bnabla\tGh\cdot\tRh^{-1} =\frac{1}{8\pi\etah}
\begin{bmatrix}
\lb(1+\frac{a^2}{6}\nabla^2\rb)\bnabla\bG \ & \ \bnabla\lb(\frac{1}{\lb|\br\rb|^3}\bXi\rb)
\end{bmatrix}.
\end{align}
and $\bG = \frac{1}{\lb|\bx\rb|}\lb(\bI +\frac{\bx\bx}{\lb|\bx\rb|^2}\rb)$ is the Oseen tensor, as shown by Lauga \cite{lauga14}. Setting $\bN = \bzero$ one obtains the result for a Newtonian fluid as shown by Stone and Samuel \cite{stone96}.

\subsection{Cylindrical swimmers}
For cylindrical swimmers, who may rotate about their axis of symmetry but translate only in the plane perpendicular to this, the stress may be written
\begin{align}
\bn\cdot\tTh = -\frac{\etah}{a}\lb[\alpha\bI_\parallel \ \  2\bXi\cdot\bI_\perp\rb],
\end{align}
where $\alpha$ is a dimensionless constant (which is singular in $\Re$). We use the $\parallel$ subscript to denote in-plane components while $\perp$ denotes the out of plane component. Integrating over the perimeter we may obtain the mobilities per unit length $\bRh_{FU}^{-1} = (2\pi\etah \alpha)^{-1}\bI_\parallel$ and $\bRh_{L\Omega}^{-1} = (4\pi\etah a^2)^{-1}\bI_\perp$. Combining these terms we obtain the swimming velocity
\begin{align}
\tU &= -\frac{1}{2\pi a}\int_{S_0}
\begin{bmatrix}
\bI\\
\frac{1}{a^2}\bXi^\top
\end{bmatrix}\cdot\bu^{S_0}\d S-\frac{\etah}{\eta}\int_{V_0} \bN:\bnabla\tGh\cdot\tRh^{-1}\d V,
\end{align}
where for a cylinder
\begin{align}
\bnabla\tGh\cdot\tRh^{-1} =\frac{1}{4\pi\etah}
\begin{bmatrix}
\lb(1+\frac{a^2}{4}\nabla^2\rb)\bnabla\bG \ & \ \frac{1}{a}\bnabla\lb(\frac{1}{r}\bXi\rb)
\end{bmatrix},
\end{align}
and $\bG = -\ln(r/a)\bI+\frac{\bx\bx}{r^2}$ is the Oseen tensor in 2D. Notice that the result is independent of $\alpha$ and thus the singular nature of the resistance problem is avoided.

\subsection{Planar swimmers}
Here we consider a two-dimensional planar swimmer (sheet), which may have different prescribed velocities, $\bu^{S_1}$ and $\bu^{S_2}$, on each side ($S_0=S_1\cup S_2$), unequally spaced between two rigid surfaces. We restrict the sheet to in-plane translation and so look only at the two-dimensional resistance problem, which is simply shear flow hence
\begin{align}
\bn\cdot\tTh = -\frac{\etah}{h}\bI_\para, \quad\quad \bnabla\tGh=-\frac{1}{h}\bn\bI_\para,
\end{align}
where now $\bI_\para$ is two-dimensional. If the distances between the swimmer and the two walls are $h_1$ and $h_2$, we have then as the only non-zero mobility per unit area $\bR_{FU}^{-1}=\frac{h_1h_2}{\etah (h_1+h_2)}\bI_\para$. Keeping in mind that the unit normal $\bn$ and distance to the wall $h$ changes depending on the side we combine with the above to obtain
\begin{align}
\bU &= \frac{1}{h_1+h_2}\Bigg(-h_2\lb<\bu^{S_1}\rb>-h_1\lb<\bu^{S_2}\rb>+\frac{1}{\eta}\lb[\int_0^{h_1}h_2\lb<\bN\cdot\bn\rb>\d x_n+\int_{0}^{h_2}h_1\lb<\bN\cdot\bn\rb>\d x_n\rb]\Bigg).
\end{align}
Here again, the angle brackets denote a surface average and $\d x_n = \d\bx\cdot\bn$. In an unbounded fluid we simply take $h_1=h_2\rightarrow \infty$ to obtain
\begin{align}\label{sheetvelocity}
\bU = -\lb<\bu^{S_0}\rb>+  \frac{1}{\eta}\langle\int_0^{\infty}\bN\cdot\bn\d x_n\rangle,
\end{align}
where the average is over both sides of the sheet.

\section{Generalized Sheet in a viscoelastic fluid}
Given the simple form of the integral theorem for swimming of two-dimensional objects we look to categorize the motion of a generally deforming flat sheet. We consider motions that are periodic in time and in space
\begin{align}
\Delta \br(\br_0,t) = A\sum_{n,p}\bc_{n,p}e^{in\bk\cdot\br_0}e^{-ip\omega t},
\end{align}
where $A$ is the characteristic amplitude of deformation, $\bk$ is the wave vector, $\br_0$ is a reference point on a flat plane with normal $\bn$ and $\bn\cdot\bk = 0$. The system is invariant in the direction $\bn\times\bk$. The Fourier coefficients, $\bc_{n,p} = a_{n,p}(\bk/\lb|\bk\rb|)+b_{n,p}\bn$, include both transverse and longitudinal deformations.
The velocity field boundary condition on the sheet is hence
\begin{align}
\bu^S = \frac{\partial \Delta \br}{\partial t} = -\omega A\sum_{n,p} ip\bc_{n,p}e^{in\bk\cdot\br_0-ip\omega t}.
\end{align}
We wish to describe this motion on the reference surface ($S_0$) and so expand in powers of $\Delta\br$
\begin{align}
\bu^{S_0} &= \bu^S -\Delta\br\cdot\lb.\bnabla\bu\rb|_{\br_0}+...\nonumber\\
&= -\omega A\sum_{n,p} ip\bc_{n,p}e^{in\bk\cdot\br_0-ip\omega t}-A\sum_{n,p}e^{in\bk\cdot\br_0-ip\omega t}\bc_{n,p}\cdot\lb.\bnabla\bu\rb|_{\br_0}+...
\end{align}
This expansion necessitates obtaining gradients of the flow field $\bu$ unless there are no transverse deformations, $\bc_{n,p}\cdot\bn = 0$. To determine the swimming motion when transverse deformations are present, one may posit a regular pertubation expansion of all fields in powers of $\epsilon \equiv A\lvert\bk\rvert \ll 1$. Using the generalized boundary conditions above, one can obtain the leading order (Newtonian) flow field, $\bu_1$, using classical methods discussed elsewhere \cite{taylor51}. By the integral formula \eqref{sheetvelocity}, the time-averaged (steady) swimming velocity in a Newtonian fluid is then immediately found to be
\begin{align}\label{NewtonU}
\overline{\bU}^N = -\lb<\overline{\bu^{S_0}}\rb> = \omega A^2\bk\sum_{p}\hat{U}^N_2(p)+\fO(\omega k^2A^3),
\end{align}
where each frequency component
\begin{align}
\hat{U}_2^N(p) = \sum_n np\lb[a_{n,p}a_{n,p}^\dagger-b_{n,p}b_{n,p}^\dagger\rb].
\end{align}
We see that all modes are decoupled. For traveling waves in the $\bk$ direction $n=p$ (or $n=-p$ for waves in the $-\bk$ direction) and we arrive at known results for a general swimming sheet with both transverse and longitudinal waves \cite{blake71b, felderhof14, elfring15} where
\begin{align}
\hat{U}_2^N(p) = p^2\lb[a_{p,p}a_{p,p}^\dagger-b_{p,p}b_{p,p}^\dagger\rb].
\end{align}
Note that the compressional and transverse waves lead to oppositely signed motion even though both waves are traveling in the same direction and so there is no net motion, at $\fO(\epsilon^2)$, if $\sum_p p^2\lb[a_{p,p}a_{p,p}^\dagger-b_{p,p}b_{p}^\dagger\rb]=0$.

Now to study the effects on viscoelasticity due to this general boundary deformation we must select a viscoelastic constitutive equation. In general, under a small-amplitude expansion we have at the first two orders
\begin{align}\label{linear}
\btau_1^{(p)}&=\eta^*(p)\bgamma_1^{(p)},\\
\btau_2^{(p)}&=\eta^*(p)\bgamma_2^{(p)}+\bN_2^{(p)}(\bu_1).
\end{align}
The leading order velocity field is Newtonian while the stress field displays both a viscous and elastic response but there is no viscoelastic effect on locomotion at this order of approximation as $\bN_1=\bzero$. Non-linearities arising from the constitutive relation enter at second order but because we are interested here only in the steady swimming speed, we need only to evaluate the mean, $\overline{\bN_2}=\bN_2^{(0)}$. We use here the Oldroyd-B constitutive equation, which in the small amplitude limit is broadly representative of nonlinear viscoelastic fluids for swimming \cite{lauga07}. The nonlinear contribution is given by the tensor
\begin{align}
\overline{\bN_2}&=\sum_p\frac{\eta_0}{ip\omega}\lb(\frac{\eta^*(p)}{\eta_0}-1\rb)\lb[\lb(\bnabla \bu_1^\text{(-p)}\rb)^\top\cdot\bgammad_1^{(p)}+\bgammad_1^{(p)}\cdot\bnabla\bu_1^{(-p)}-\bu_1^{(-p)}\cdot\bnabla\bgammad_1^{(p)}\rb].\nonumber
\end{align}

Armed with the solution of the leading order Newtonian flow field $\bu_1$, only straightforward integration, via \eqref{sheetvelocity}, is then needed to obtain the swimming velocity for a generally deforming flat sheet in a (Oldroyd-B) viscoelastic fluid, to leading order
\begin{align}\label{complexU}
\overline{\bU} = \omega A^2\bk\sum_{p}\frac{\eta^*(p)}{\eta_0}\hat{U}^N_2(p).
\end{align}
We see from the above equation that the contribution of each mode is rescaled by the real part of the dimensionless complex viscosity. This factor is always less than or equal to one,
\begin{align}
\Re\lb[\frac{\eta^*(p)}{\eta_0}\rb]=\frac{1+p^2\De^2\beta}{1+p^2\De^2}\le 1,
\end{align}
because the retardation time, $\lambda_2$, is smaller than the relaxation time, $\lambda_1$, in viscoelastic fluids, $\beta=\lambda_2/\lambda_1 < 1$. This means that the swimming speed due to any individual temporal mode is slower than in a Newtonian fluid. The Deborah number $\De = \omega\lambda_1$ characterizes the response of the fluid, when the time scale of actuation is much longer than that of the relaxation of the fluid, $\De\rightarrow 0$, and we recover the Newtonian swimming speed. Physically, the thrust (by way of equation \eqref{Usmall}) is reduced by a factor equal to the frequency dependent viscosity, meaning the higher frequencies are increasingly damped. Now because each individual mode is less effective as a means of propulsion,
\begin{align}
\Re\lb[\frac{\eta^*(p)}{\eta_0}\rb]\lb|\hat{U}^N_2(p)\rb|\le \lb|\hat{U}^N_2(p)\rb|,
\end{align}
slower swimming in a viscoelastic fluid is guaranteed if $\hat{U}^N_2(p)$ does not change sign $\forall p$, an example being a sheet passing only unidirectional transverse waves. However, swimming need not be slower in general because $\hat{U}^N_2(p)$ can certainly change sign in more general deformations. In particular, one may note that unidirectional transverse and longitudinal waves lead to motion in opposite directions and therefore may result in faster swimming in a viscoelastic fluid. 

A simple example of faster swimming is found by recalling that in a Newtonian fluid the swimming speed (at $\fO(\epsilon^2)$) is zero when a sheet is passing unidirectional transverse and compressional deformation waves if $\sum_p p^2\lb[a_{p,p}a_{p,p}^\dagger-b_{p,p}b_{p,p}^\dagger\rb]=0$. In a viscoelastic fluid the same actuation may lead to net motion because the coefficients are rescaled by a factor that diminishes for higher modes. For example, if we take two modes, $q$ and $m$, where $q^2b_{q,q}b_{q,q}^\dagger=m^2a_{m,m}a_{m,m}^\dagger$ and all other modes zero, in a Newtonian fluid the thrust due to the two modes cancels one another and therefore there is no net motion at $\fO(\epsilon^2)$, $\overline{\bU}^N=\omega A^2\bk\sum_{p}\hat{U}^N_2(p)=\bzero$. In a viscoelastic fluid however, the higher frequency term is damped by a larger factor and hence the lower frequency component generates a larger thrust leading to the swimming speed
\begin{align}
\overline{\bU}&=\omega A^2\bk\sum_{p}\frac{\eta^*(p)}{\eta_0}\hat{U}^N_2(p),\nonumber\\
&=\omega A^2\bk m^2a_{m,m}a_{m,m}^\dagger\frac{2(q^2-m^2)(1-\beta)\De^2}{(1+m^2\De^2)(1+q^2\De^2)},\label{2elastic}
\end{align}
which is maximum when $\De = 1/\sqrt{mq}$ and decays quadratically as $\De\rightarrow 0$. If $q^2>m^2$ then the swimmer moves in the direction of $\bk$ as the thrust due to the compressional waves is dominant, if $q^2<m^2$ the opposite is true, while if $m=q$ there is only a single mode and the swimming speed is zero.

Because higher frequencies are damped by larger factors, swimmers may go faster in a viscoelastic fluid compared to a Newtonian fluid, as in the previous example, depending on the amplitudes of the modes and in general, the Deborah number. Lauga and Riley explore this effect with oppositely traveling transverse waves (for which $\hat{U}_2^N(p) = -p^2\lb[b_{p,p}b_{p,p}^\dagger-b_{-p,p}b_{-p,p}^\dagger\rb]$ can change sign) finding the necessary conditions for two such waves to yield faster swimming \cite{riley15}. Generally a swimmer may go faster, slower, or even stop in a viscoelastic fluid provided there are multiple modes and $\hat{U}^N_2(p)$ changes sign.

It may seem obvious that the complex viscosity, measureable by the linear response \eqref{linear}, determines the change in swimming speed due to a viscoelastic fluid, but if the fluid were only linearly viscoelastic there would be absolutely no change in the swimming speed from that of a Newtonian fluid. It is the nonlinear response that leads to change in the swimming speed, not the linear response, the tensor $\bN_2$ happens to lead to a rescaling of the Newtonian solution by the viscous modulus and that result is remarkably robust. The same result holds for a sheet swimming near a wall and two-dimensional pumping \cite{elfring15b}, as well as helical swimming, both in an unbound fluid and near wall \cite{li15}. The question one might ask is why should each mode, individually, be slower? In other words, why should the thrust generated by any single component of a deforming boundary condition necessarily be diminished in a viscoelastic fluid? Ultimately we find that the nonlinear non-Newtonian response of the fluid is of the same form as the motion due to the deforming boundary in a Newtonian fluid, but oppositely signed, namely
\begin{align}
\frac{1}{\eta_0}\int_0^{\infty}\lb<\overline{\bN_2}[\bu_1]\cdot\bn\rb>\d x_n &=\sum_p\lb[\frac{\eta^*(p)}{\eta_0}-1\rb]\lb<\br_1^{(-p)}\cdot\bnabla\bu_1^{(p)}\rb>,\nonumber\\
&=-\omega A^2\bk \sum_{p}\De\frac{G^*(p)}{\eta_0 \omega}\hat{U}_2^N(p).
\end{align}
Here we have written the response in terms of the elastic modulus $G^* = -ip\omega \eta^*$ ,whose real part is
\begin{align}
\text{Re}\lb[\frac{G^*(p)}{\eta_0 \omega}\rb]=\frac{p^2\De(1-\beta)}{1+p^2\De^2},
\end{align}
and so when summed with the Newtonian contribution the result for each mode is always slower. We will refer to this as an elastic reponse because the contribution of each mode is scaled by the (real part of the) elastic modulus, which goes to zero as $\De\rightarrow 0$.

This begs the question of whether one may elicit an elastic response from the fluid by prescribing a velocity boundary condition with no associated deformation? We find that if we directly prescribe a slip velocity $\bu^{S} = \bu^{S_0}$, without deforming the body, $\Delta \br=\bzero$, for example by modifying surface chemistry \cite{golestanian07}, the results can be markedly different. If we prescribe a general time-periodic slip velocity on a \textit{flat} sheet,
\begin{align}
\bu^{S_0}  = -\omega A\sum_{n,p} ip\bc_{n,p}e^{in\bk\cdot\br_0-ip\omega t},
\end{align}
to leading order this is the same boundary condition as prescribed above for a deforming sheet, but because here there is no associated deformation $\lb<\overline{\bu^{S_0}}\rb>=\bzero$ at all orders and the contribution to the swimming speed is entirely due to viscoelastic part of the thrust, namely
\begin{align}
\overline{\bU} = -\omega A^2\bk \sum_{p}\De\frac{G^*(p)}{\eta_0 \omega}\hat{U}_2^N(p).\label{slipelastic}
\end{align}
We see a quadratic decay to zero as $\De\rightarrow 0$ because, as constructed, there cannot be any net motion for these boundary conditions in a Newtonian fluid. For large Deborah numbers the swimming speed approachs the limit $\overline{\bU}\rightarrow -(1-\beta)\overline{\bU}^N$ as $\De\rightarrow \infty$ where $\overline{\bU}^N$ refers to the swimming speed of a \textit{deforming} sheet in a Newtonian fluid as given by \eqref{NewtonU}. Analogously, the response is larger for higher modes. We also note that we were able to construct a response of a similar form (see \eqref{2elastic}) with traveling deformation waves by superposing both transverse and longitudinal waves whose thrust is in opposition. If the thrust of these modes cancels exactly in a Newtonian fluid, as it did in our example, the swimming speed is determined by an elastic response, while if the two terms do not cancel exactly, there is an additional contribution to the swimming speed that scales as $\Re[\eta^*/\eta_0]$.

We often see this form of elastic mediated response, $U\propto \Re[G^*/\eta_0 \omega]$, when no net motion would occur if the fluid were Newtonian. A similar response was found by Lauga for a squirmer with a slip velocity prescribed so that the swimming speed in a Newtonian fluid is zero \cite{lauga09}. Pak \textit{et al.} also found the swimming velocity $\propto \De(1-\beta)/(1+\De^2)$ for two co-rotating spheres that propel due to the symmetry-breaking axial flows generated by viscoelastic hoop stresses \cite{pak12}, and also similarly found by Yazdi and Ardekani for a reciprocal squirmer near a wall \cite{yazdi14}. One also observes an elastic mediated synchronization of a system of two swimmers in a viscoelastic fluid that conversely displays no relative phase evolution in a Newtonian fluid \cite{elfring10,elfring15b}. A similar response is also noted for force generation by the flapping motion of a rigid rod \cite{normand08}. Naturally one might prescribe a slip velocity, with or without deformation, that would lead to swimming both in a Newtonian fluid and in a non-Newtonian fluid and therefore the swimming velocity could contain terms that vary as both the viscous and elastic modulus of the fluid, but if these boundary conditions are meant to emulate a deforming body then the result may not be physically sensible.

\section{Conclusion}
In this paper, we used a reciprocal theorem formulation to study the motion of swimming bodies in non-Newtonian fluids. We first used this formulation to show that generally, for small amplitude deformations, the leading order change in swimming speed is due to a modification of thrust generated by the swimmer in complex fluids. We then used this theory to study the effect of viscoelastic stresses on the swimming velocity of a sheet undergoing both general transverse and longitudinal deformations. We showed that it is possible to swim faster or slower, depending on the swimming gait, not only with oppositely traveling transverse waves but also with unidirectional traveling waves provided both transverse and longitudinal deformation modes are present and then demonstrate why and how this occurs, thereby extending the results of Riley and Lauga to this case \cite{riley15}. What becomes clear, even in this simplified model, is that it may be quite deceptive to draw any conclusion about whether one swimmer should go faster or slower based only on observations of a different (simpler) swimmer, because of the unequal damping of higher and lower frequency deformation modes in viscoelastic fluids. In particular, it is only with multiple (small amplitude) deformation modes that a swimming speed enhancement can be realized in viscoelastic fluids. Additionally, we found that while swimming by deformation elicits a particular response in which all swimming modes are individually diminished in a viscoelastic fluid, one may prescribe slip velocities that lead to a completely different (elastic) response in which all modes, individually, lead to a speed increase.

It remains to be seen how these results extend to finite bodies and whether a swimming gait may be decomposed into components that yield a viscous or an elastic response from the fluid. For example, numerical results indicate that even finite filaments passing sinusoidal deformation waves do not experience a speed increase in viscoelastic fluids, yet when a front-back asymmetry in amplitude (resembling flapping) is introduced an elastic response is excited that leads to a speed increase \cite{teran10}, and when that amplitude asymmetry is reversed a speed decrease is conversely observed \cite{thomases14}. Experimental work by Shen and Arratia \cite{shen11}, with the nematode \textit{C. elegans}, revealed a trend of decreasing speed with the increase in $\De$. According to Sznitmann et al.\cite{sznitman10,sznitman10b}, \textit{C. elegans} passes a transverse deformation consisting of opposite traveling waves with identical frequencies and our results indicate that this would lead to slower speeds, at small amplitudes, due to the lack of multiple modes required for enhancement. The front-back amplitude asymmetry of the swimmer also has an effect\cite{thomases14}, but such effects may be minimal for swimmers much longer than a typical wavelength\cite{liu11, spagnolie13}.

The picture developed here is only valid at small amplitudes and it is not clear in what ways large amplitude motions change the response of the fluid. There is indication that a single-mode swimming sheet is always slower, even for large amplitude deformations, as demonstrated both analytically \cite{elfring15b} and numerically \cite{li15b}, but the opposite trend was found for helical swimmers, which are slower at small amplitude but were observed, both experimentally \cite{liu11} and numerically \cite{spagnolie13}, to swim faster at large amplitude. For large amplitude deformations one should also consider the effects of finite extension as microorganisms typically reside in regimes where the Deborah number is not small\cite{lauga07,lauga09}. Finally, we note that the results presented here are limited to swimmers with a prescribed gait and the picture is further complicated when fluid stresses lead to changes in the gait itself \cite{curtis13, riley14}.

\acknowledgements
The authors thank Eric Lauga and Arthur Evans for many fruitful discussions and acknowledge funding from the Natural Science and Engineering Research Council of Canada (NSERC). The authors also thank anonymous referees for their helpful suggestions.

\appendix

\bibliography{swimming}
\end{document}